\documentstyle[epsfig,preprint,aps]{revtex}
\tightenlines 
\def\bbbone{{\mathchoice {\rm 1\mskip-4mu l} {\rm 1\mskip-4mu l}
{\rm 1\mskip-4.5mu l} {\rm 1\mskip-5mu l}}}
\def\bbbc{{\mathchoice {\setbox0=\hbox{$\displaystyle\rm C$}\hbox{\hbox
to0pt{\kern0.4\wd0\vrule height0.9\ht0\hss}\box0}}
{\setbox0=\hbox{$\textstyle\rm C$}\hbox{\hbox
to0pt{\kern0.4\wd0\vrule height0.9\ht0\hss}\box0}}
{\setbox0=\hbox{$\scriptstyle\rm C$}\hbox{\hbox
to0pt{\kern0.4\wd0\vrule height0.9\ht0\hss}\box0}}
{\setbox0=\hbox{$\scriptscriptstyle\rm C$}\hbox{\hbox
to0pt{\kern0.4\wd0\vrule height0.9\ht0\hss}\box0}}}}

\begin{document}

\draft
\title{
Quantum Projector Method on Curved Manifolds}
\author{V. Melik-Alaverdian$^a$, G. Ortiz$^b$, and N. E. Bonesteel$^c$}
\address{$^a$
Department of Physics, University of Rhode Island, Kingston, RI 02881}
\address{$^b$
Theoretical Division, Los Alamos National Laboratory, P.O. Box 1663, 
Los Alamos, NM 87545}
\address{$^c$
National High Magnetic Field Laboratory and Department of Physics,
Florida State University, Tallahassee, FL 32306-4005}

\date{\today}
\maketitle

\begin{abstract}
A generalized stochastic method for projecting out the ground state of
the quantum many-body Schr\"odinger equation on curved manifolds is
introduced. This random-walk method is of wide applicability to any
second order differential equation (first order in time), in any
spatial dimension. The technique reduces to determining the proper
``quantum corrections'' for the Euclidean short-time propagator that
is used to build up their path-integral Monte Carlo solutions. For
particles with Fermi statistics the ``Fixed-Phase'' constraint (which
amounts to fixing the phase of the many-body state) allows one to
obtain stable, albeit approximate, solutions with a variational
property. We illustrate the method by applying it to the problem of an
electron moving on the surface of a sphere in the presence of a Dirac
magnetic monopole. 
\end{abstract}



\columnseprule 0pt

\narrowtext

\section{Introduction}
\label{section1}

The importance and difficulty of solving models of interacting quantum 
particles is hard to overstate. It is well known that the correlated 
motion of those particles gives rise to a wide variety of physical 
phenomena at different length and time scales, spanning disciplines 
like chemistry, condensed matter, nuclear, and high energy physics. 
Novel complex structures can emerge as a consequence of the competing
multiple-length scales in the problem. Nonetheless, only a reduced  set
of interacting problems admits exact closed form solutions
\cite{mattis} and the  use of numerical techniques becomes essential if
one is looking for  accurate solutions not subjected to uncontrolled
approximations. Among  those techniques, the statistical methods
\cite{negele} offer the potential to study  systems with large number
of degrees of freedom, reducing the  computational complexity from
exponential to polynomial growth. This  scaling behavior is
particularly relevant when one recognizes that  most of the interesting
phenomena in many-body physics occurs in the  thermodynamic limit
\cite{anderson}. Unfortunately, for fermions (i.e. quantum  particles
obeying Fermi statistics) the sign problem plagues all  useful
stochastic algorithms and causes the variance of computed  results to
increase exponentially with increasing number of fermions \cite{ceper}. 

On the other hand, the growing interest in physical systems whose 
state functions are defined on a general metric space makes the 
quantum mechanics of interacting particles in curved manifolds no 
longer a mere intellectual exercise, but one with very practical 
consequences. Perhaps the most well-known examples can be found in 
cosmology (e.g., matter in strong gravitational fields, atomic 
spectroscopy as probe of space-time curvature \cite{gravitation}), but
the subject is  certainly not exclusive to this field. In condensed
matter a very  elementary case is provided by a deformed crystal. Less
well-known  ones are mesoscopic graphitic microtubules and fullerenes.
All these  physical systems are ubiquitous in nature and the crucial
role the  curvature of the manifold plays has been confirmed by
experimental  observations (e.g. spectrum of collective excitations
\cite{curve}). Therefore, the development of stable quantum methods
with polynomial complexity in Riemannian manifolds represents a real
challenge for many-body  theorists.

The present manuscript deals with the (non-relativistic) many-particle 
Schr\"odinger equation in a general metric space and its solution 
using stochastic techniques. In particular, we will show how to 
construct approximate solutions (wave functions) for systems with 
broken time-reversal symmetry (e.g. electrons in the presence of 
external electromagnetic sources) avoiding the infamous ``phase 
problem'' \cite{fp}. The main difficulty is to define a probability
measure  (semi-positive definite) which allows one to get the
complex-valued  state with no asymptotic signal-to-noise ratio decay in
Euclidean time. This translates into a problem of geometric
complexity, which is  solved approximately using constraints in the
manifold where the wave function has its support. In this way, we get
stable but approximate solutions which can be systematically
improved.  

Among the large variety of problems one can attack, we decided to 
choose the general problem of fermions in the presence of external
gauge fields to illustrate the main ideas. The effects of an external 
magnetic field on a system of electrons can be profound \cite{landau}.
The field  couples to the electron's charge and spin, modifying its
spatial  motion and lifting its spin degeneracies. The field can also
create  spatial anisotropy, effectively reducing the dimensionality of
the  system from three to two. The combination of the reduced dimension
and  the field itself is known to have novel consequences. For example,
in  a system of non-interacting electrons hopping on a square lattice,
the  field transforms the energy spectrum from the simplicity of 
trigonometric functions to the complexity of a field-dependent 
self-similar structure (Hofstadter's butterfly) whose depth 
mathematicians are still fathoming \cite{hofst}. The combination of 
the reduced dimensionality, strong particle interactions and the field 
itself is known to have novel consequences, like the formation of 
isotropic fractional quantum Hall fluids \cite{QHE}, which are  
incompressible states of the two-dimensional homogeneous Coulomb gas.

The projector (zero temperature) method we will introduce uses 
random-walks to solve a general multidimensional partial differential 
equation second order in space coordinates and first order in time. 
Whenever mention is made of a random-walk we mean a Markov chain that 
is defined as a sequence ${\cal R}_1,{\cal R}_2, \cdots, {\cal R}_K $ 
of $K$ random variables that take values in configuration space, i.e. 
the space of particle positions. As usual, what characterizes a 
random-walk is its initial probability distribution and a conditional 
probability that dictates the transition from ${\cal R}_i$ to ${\cal 
R}_{i+1}$. This transition probability is non-unique and 
discretization dependent \cite{tira}. Among all the possible choices 
we will require a prepoint discretization of the transition 
probability (short-time propagator) because we will use Monte Carlo 
methods to generate the walkers. 

The paper is organized as follows. In Section \ref{section2} we 
present the formulation of the general problem of fermions in curved 
manifolds. In particular, for illustration purposes and to fix 
notation, we develop the formalism for spin-$\frac{1}{2}$ particles in 
the presence of an external electromagnetic potential. Then, we show 
how to project out the lowest energy state of a given symmetry  in a
manifold with curvature, and discuss the resulting Fokker-Planck 
equations for various distribution functions. Once the problem is 
precisely defined we develop, in Section \ref{section3}, path-integral 
solutions to those multidimensional differential equations, and 
give an interpretation of the emergent ``quantum corrections'' in the 
Euclidean action. The path-integral solutions are evaluated using 
Monte Carlo techniques in Section \ref{section4}. There, we provide  an
stable step by step practical algorithm which emphasizes the subtle 
changes (with respect to the standard Diffusion Monte Carlo (DMC)
technique)  due to the metric of the manifold. In Section
\ref{section5} we apply such computational implementation to the
problem of an  electron moving on the surface of a sphere in the
presence of a Dirac  monopole. Finally, Section \ref{section6}
summarizes the main  findings and discusses the relevance of the
stochastic method as  applied to the physics of quantum Hall fluids.  

\section{Fermions on Riemannian manifolds}
\label{section2}

{\it Notation.} 
Consider a differentiable manifold $\cal M$ of  dimension $d$ (e.g.,
for the two-sphere S$^2$, $d$=2) with coordinates  ${\bf r}_i =
(x_i^1,\ldots,x_i^d)$ defined on it. If ${\cal M}$ is a  Riemannian
manifold, then it is a metric space, with metric tensor  $g^{\mu
\nu}({\bf r}_i)=g^{\mu \nu}(i)$, such that the distance $d s$  between
two points in  ${\cal M}$ is $ds^2 = g_{\mu \nu}(i) \ dx_i^{\mu}
dx_i^{\nu}$ in the  usual way \cite{landauctf}. The metric tensor is
positive definite and symmetric  $g^{\mu \nu}=g^{\nu \mu}$ (as we will
see, this condition is important  to define a probability density
distribution), and is a function of  the coordinates ${\bf r}_i$ with
the property $g_{\mu \gamma}  g^{\gamma \nu} = \delta^{\nu}_{\mu}$. Let
us consider the coordinate  transformation $h$: $x_i^{\mu} = h^{\mu}
(x'^1_i,\ldots,x'^d_i)$.  Then, a generic second order contravariant
($T^{\mu \nu}$) and  covariant tensor ($T_{\mu \nu}$) transform as 
\begin{equation}
T^{\mu \nu} = \frac{\partial x^{\mu}_i}{\partial x'^{\alpha}_i} 
\frac{\partial x^{\nu}_i}{\partial x'^{\beta}_i} \ T'^{\alpha \beta} 
\;\;\; , \;\;\; T_{\mu \nu} = \frac{\partial x'^{\alpha}_i}{\partial 
x^{\mu}_i} \frac{\partial x'^{\beta}_i}{\partial x^{\nu}_i} 
T'_{\alpha \beta} \ ,
\end{equation}
respectively. Throughout the paper Einstein's summation convention for 
repeated indices is assumed ($\mu,\nu=1,\ldots,d$).

{\it Formulation of the problem.} 
In this article we will be concerned with finite interacting fermion 
systems in the presence of an external electromagnetic potential 
$a_{\mu}({\bf r}_i) = a_{\mu}(i) = ({\bf A}(i),\phi(i)=0)$ (${\bf B} = 
{\bf \nabla \wedge A}$ represents a uniform field, ${\bf A}$ and 
$\phi$ are the vector and scalar potentials, respectively) whose 
Hamiltonian for motion on the manifold, in the coordinate 
representation, is given by 
\begin{equation}
\widehat{\rm I\!H} = \widehat{\rm I\!H}_0 + \widehat{V}(\{ {\bf r}_i 
\},\{s_i\})
\end{equation}
with 
\begin{equation}
\widehat{\rm I\!H}_0 = - D \mbox{\boldmath $\Delta$} + i \frac{e 
\hbar}{2 m^* c} \sum_{i=1}^{N} \left[ 2 a^{\mu} (i)  
\partial_{\mu} + g^{-1/2}(i) \partial_{\mu} \left( 
g^{1/2}(i) a^{\mu}(i) \right) \right] + \frac{e^2}{2 m^* c^2} 
\sum_{i=1}^{N} a^{\mu}(i) a_{\mu}(i) \ ,
\end{equation} 
$\partial_{\mu} = \partial/\partial x_i^{\mu}$, and $\mbox{\boldmath 
$\Delta$} = \sum_{i=1}^{N} \Delta (i)$, where 
\begin{equation}
\Delta = g^{-1/2} \partial_{\mu} \left( g^{\mu \nu}g^{1/2} 
\partial_{\nu} \right) 
\end{equation} 
is the covariant Laplace-Beltrami operator and $\widehat{V}$ is a 
potential energy operator. Notice that we use the 
conventional notation where the transformation between 
different forms of a given tensor is achieved by using the metric 
tensor (e.g., $a^{\mu} = g^{\mu \nu} a_{\nu}$, $ a_{\mu} = 
g_{\mu \nu} a^{\nu}$), and $g^{1/2} = \sqrt{{\rm det} \ g_{\mu\nu}}$. 
This Hamiltonian characterizes the dynamics of $N$ non-relativistic 
indistinguishable particles of mass $m^*$, charge $e$ and spin 
$s_i=\frac{1}{2}$ in a curved space with metric tensor $g^{\mu \nu}$, 
and $D = \hbar^2/2m^*$. We have assumed that the quantum Hamiltonian 
$\widehat{\rm I\!H}$ in curved space has the same form as in flat 
space (this amounts to a particular operator ordering prescription.) 

Given the previous ordering, one can rewrite the Hamiltonian above
in terms of the generalized (hermitian) canonical momentum ${\bf 
p}_{\mu}= - i \hbar (\partial_{\mu} + \frac{1}{2} 
\partial_{\mu}(\ln g^{1/2}))$
\begin{equation}
\widehat{\rm I\!H} = {1 \over 2 m^*} \sum_{i=1}^{N} g^{-1/4}(i) \ {\bf 
\Pi}_{\mu} \ g^{1/2}(i) \ g^{\mu \nu}(i) {\bf \Pi}_{\nu} \ g^{-1/4}(i) 
+ \widehat{V} (\{ {\bf r}_i \},\{s_i \}) \ ,
\label{eq:hami}
\end{equation}
where the kinetic momentum ${\bf \Pi}_{\mu} = {\bf p}_{\mu} - {e 
\over c} a_{\mu}$. The first term in Eq. \ref{eq:hami}  
represents the kinetic energy of the system and is the 
non-relativistic approximation to the Dirac operator. $\widehat{V}$ 
includes the sum of one and 
two-body local interaction terms (and background potential in the 
case of a charge neutral system) and Zeeman contribution. The 
potentials are assumed to be finite almost everywhere and can only be
singular at coincident points (${\bf r}_i = {\bf r}_j$, $\forall \ i 
\neq j$) 

We are interested in the stationary solutions of the resulting 
multidimensional Schr\"o\-dinger equation 
\begin{equation}
i \hbar \ \partial_t | \Psi \rangle  = \widehat{\rm I\!H} \ | \Psi 
\rangle \ ,
\label{sch}
\end{equation}
and will restrict ourselves to Hamiltonians $\widehat{\rm I\!H}$ which 
are time-translation invariant. In the usual space-spin formalism the 
$N$-fermion states characterizing the system, $\langle X|\Psi \rangle 
= \Psi(X)$, and all its 
first derivatives belong to the Hilbert space of antisymmetric (with 
respect to identical particle $({\bf r}_i,s_i)$-exchanges) 
square-integrable functions ${\cal H}_N = {\cal L}^2({\cal M}^{N}) 
\otimes {{\mathchoice {\setbox0=\hbox{$\displaystyle\rm C$}\hbox{\hbox 
to0pt{\kern0.4\wd0\vrule height0.9\ht0\hss}\box0}}
{\setbox0=\hbox{$\textstyle\rm C$}\hbox{\hbox
to0pt{\kern0.4\wd0\vrule height0.9\ht0\hss}\box0}}
{\setbox0=\hbox{$\scriptstyle\rm C$}\hbox{\hbox
to0pt{\kern0.4\wd0\vrule height0.9\ht0\hss}\box0}}
{\setbox0=\hbox{$\scriptscriptstyle\rm C$}\hbox{\hbox 
to0pt{\kern0.4\wd0\vrule height0.9\ht0\hss}\box0}}}}^{2 {N}}$, 
defined as
\begin{equation} 
{\cal H}_N \:=\: \left\{ \Psi \: \mid \: \widehat{P}_{ij} \Psi\:=\: 
- \Psi \;,\;{\rm and}\; \| \Psi \| = \sqrt{\langle \Psi | \Psi 
\rangle} <\: \infty \right\} \ ,
\end{equation} 
where $X = ({\cal R}, {\it \Sigma})$ (${\cal R} = ({\bf 
r}_1,\ldots,{\bf r}_N)$ and ${\it \Sigma} = 
(\sigma_1,\ldots,\sigma_N)$ are discrete spin variables) 
and $\widehat{P}_{ij}$ represents the permutation of 
the pairs $({\bf r}_i, \sigma_i)$ and $({\bf r}_j, \sigma_j)$. 
${\cal M}^{N}$ is the Cartesian product manifold of dimension $d N$. 

Since the system Hamiltonian can be written as $\widehat{\rm I\!H}\:=\:
\widehat{\rm I\!H}_{\cal R} ({\cal R})\:+\:\widehat{\rm I\!H}_S 
({\it \Sigma})$, the last term representing the Zeeman coupling, the 
many-body wave function $\Psi ({\cal R},{\it \Sigma})$ can 
be expressed as a tensor product of a coordinate and a spin 
function (or a linear combination of such products), 
\begin{equation}
\Psi ({\cal R},{\it \Sigma}) \:=\: \Phi ({\cal R})\: \otimes \: \Xi (
{\it \Sigma}) \;.
\label{mc2}
\end{equation}  
We want to construct $N$-fermion eigenstates of $\widehat{\rm I\!H}$, 
$\Psi$ , that are also eigenfunctions of the total spin $S^2$ ($S = 
\sum_{i=1}^N s_i$),
\begin{equation}
S^2 \: \Psi (X) \:=\:\hbar^2\: s (s+1)\:\Psi (X) \; , 
\end{equation}  
and this is always possible since $\left[\widehat{\rm I\!H}, S^2 
\right]$ = 0. Thus, the configuration part $\Phi ({\cal R})$ must have 
the right symmetry in order to account for the Pauli principle. It
turns out that a coordinate state $\Phi ({\bf r}_1, \ldots, {\bf r}_k, 
{\bf r}_{k+1}, \ldots, {\bf r}_N)$ which is symmetrized according to 
the Young scheme \cite{hammer} and has total spin $s = {N \over 
2} \:-\: k $ will be antisymmetric in the variables ${\bf r}_1, 
\ldots, {\bf r}_k$, and antisymmetric in the variables ${\bf r}_{k+1}, 
\ldots, {\bf r}_N$. Moreover, $\Phi$ possesses the property of Fock's 
cyclic symmetry, 
\begin{equation}
\left( \bbbone \:-\: \sum_{j=k+1}^N \; \widehat{P}_{kj} \right)\: \Phi 
\:=\: 0 \;,
\end{equation}
where, in this case, $\widehat{P}_{kj}$ refers to the transposition of 
particle coordinates ${\bf r}_k$ and ${\bf r}_j$. This last condition 
is a very useful one for testing the symmetry of a given coordinate 
function. 

{\it Quantum projection on curved manifolds.} 
For a given total spin $s$ we are thus left with the task of solving 
the stationary many-body Schr\"odinger equation $\widehat{{\rm 
I\!H}}_{\cal R} \Phi ({\cal R}) = E \Phi ({\cal R})$, where $\Phi 
({\cal R}) = \langle {\cal R} | \Phi \rangle$ satisfies 
the symmetry constraint discussed above. In particular, we are 
interested in the zero temperature properties of this 
quantum system, i.e. its ground state properties. To this end, we 
study the Euclidean time evolution of the state $\Phi$, i.e. we 
analytically continue Eq. \ref{sch} to imaginary time (Wick 
rotation, $t \rightarrow -i t \hbar$) 
\begin{equation}
- \partial_t \Phi  = \left[ \widehat{\rm I\!H}_{\cal R} - E_T \right] 
\Phi \ ,
\label{scht}
\end{equation}
whose formal solution $\Phi(t) = \hat{\cal U}(t) \ \Phi_T = \exp [ 
-t(\widehat{\rm I\!H}_{\cal R} - E_T)] \Phi_T$ is used to determine 
the limiting distribution 
\begin{equation}
\Phi_0 \propto \lim_{t \rightarrow \infty} \Phi(t)\ ,
\end{equation}
which is the largest eigenvalue solution of the evolution operator 
$\hat{\cal U}(t)$ compatible with the 
condition $\langle \Phi_0 | \Phi_T \rangle \neq 0$, where $\Phi_T$ is 
a parent state and $E_T$ is a suitable (constant) energy that shifts 
the zero of the spectrum of $\widehat{\rm I\!H}_{\cal R}$. 

We would like to solve the multidimensional differential equation 
Eq. \ref{scht} using initial value random walks. In this way, 
starting with an initial population of walkers (whose state space is 
${\cal M}^N$) distributed according to $p({\cal R}, t=0) = \Phi_T$ 
($\Phi_T$ must be positive semi-definite), the ensemble is evolved 
by successive applications of the short (imaginary) time propagator 
$\hat{\cal U}(\tau)$ ($\tau = t/M$, and $M$ is the number of 
time slices) to obtain the limiting distribution $\Phi_0$. Then, we 
can introduce a ``pseudo partition function''
\begin{equation}
{\cal Z} = \langle \Phi_T | \ \hat{\cal U}(t) \ \Phi_T \rangle 
\end{equation}
in terms of which we can determine the ground state energy $E_0$ as
\begin{equation}
E_0 - E_T = \lim_{t \rightarrow \infty} \ -\frac{1}{t} \ln {\cal Z} 
\ .
\end{equation}
Similarly, other ground state expectation values, e.g., $\langle 
\Phi_0 | \widehat{\cal O} \ \Phi_0 \rangle$, can be obtained as 
derivatives (with respect to a coupling constant $J$) of a modified 
pseudo partition function ${\cal Z}_J$ whose evolution operator has 
a modified Hamiltonian, $\widehat{\rm I\!H}_{\cal R} + J 
\widehat{\cal O}$. 

In order to reduce statistical fluctuations in the measured quantities 
(i.e., observables) one can guide the random walk with an approximate 
wave function, $\Phi_G$, which contains as much of the essential 
physics as possible (including cusp conditions at possible 
singularities of the potential $\widehat{V}$). Then, instead of 
sampling the wave function $\Phi(t)$ one samples the distribution 
$\tilde{f}({\cal R},t) = \Phi(t) \Phi_G$ (properly normalized) with 
the initial time condition $\tilde{f}({\cal R},t=0) = \Phi_T \Phi_G$. 
Expectation values of operators $\widehat{\cal O}$ (observables)
that commute with the Hamiltonian have a particularly simple form 
for guided walkers. For instance,
\begin{equation}
\lim_{t \rightarrow \infty} \frac{\langle \Phi_T | \ \widehat{\cal O} 
\ \hat{\cal U}(t) \ \Phi_T \rangle}{\langle 
\Phi_T | \ \hat{\cal U}(t) \ \Phi_T \rangle} = 
\langle \Phi_G^{-1} \widehat{\cal O} \Phi_T \rangle_{\tilde{f}(t 
\rightarrow \infty)} \ ,
\label{mixede}
\end{equation}
where the average $\langle {\cal A} \rangle_{\tilde{f}}$ stands for
\begin{equation}
\langle {\cal A} \rangle_{\tilde{f}} = \frac{\int_{{\cal M}^{N}} 
\mbox{\boldmath $\omega$} \ \tilde{f}({\cal R},t \rightarrow \infty) 
\ {\cal A}({\cal R})}{\int_{{\cal M}^{N}} 
\mbox{\boldmath $\omega$} \ \tilde{f}({\cal R},t \rightarrow \infty)} 
\ ,
\end{equation}
$\tilde{f}({\cal R},t \rightarrow \infty)$ is the long-time stationary 
probability of the system, and the (invariant) volume element 
$\mbox{\boldmath $\omega$}$ is given by the $d N$-form
\begin{equation}
\mbox{\boldmath $\omega$} = \left[ \prod_{i=1}^N g^{1/2}(i) \right] \ 
{\rm d}x_1^1 \wedge \cdots {\rm d}x_1^d \wedge \cdots {\rm d}x_{ N}^d 
\ .
\end{equation}
Remember that in a general metric space the resolution of the 
identity operator with respect to the spectral family of the 
position operator is 
\begin{equation}
\bbbone = \int_{{\cal M}^{N}} \mbox{\boldmath $\omega$} \ 
| {\cal R} \rangle \langle {\cal R} | \ .
\end{equation}
It is important to stress that $\Phi_T$ and the guiding function 
$\Phi_G$ can, in principle, be different functions, although most of 
the practical calculations use the same function. It turns out that 
this importance sampling procedure is decisive to get sensible results 
when the potential $\widehat{V}$ presents some singularities. 

Notice, however, that the quantum Hamiltonian $\widehat{{\rm I\!H}}$ 
breaks explicitly time-reversal symmetry, meaning that in general 
$\Phi$ will be a complex-valued function. Even if $\Phi$ were 
real-valued, because it represents a fermion wave function it can 
acquire positive and negative values (the case where $\Xi({\it 
\Sigma})$ is totally antisymmetric being the exception). Then, it is 
clear that we cannot in principle interpret $\Phi$ or $\tilde{f}$ as a 
probability density. 

For reasons that will become clear later we will be interested in 
sampling the probability density $\bar{f}({\cal R},t) = 
|\tilde{f}({\cal R},t)|$. The generalized diffusion equation in curved 
space for the importance-sampled function $\bar{f}$ can be derived 
directly from Eq.~\ref{scht} with the result
\begin{equation}
\partial_{t} \bar{f} = D \sum_{i=1}^{N} \left [ g^{-1/2}(i) 
\partial_{\mu} \left( g^{\mu \nu}(i) g^{1/2}(i) ( \partial_{\nu} 
\bar{f} - \bar{f} \ F_{\nu} ) \right)\right] - (E_L-E_T) \bar{f} \ ,
\end{equation}
where the drift velocity $F_{\nu}({\cal R}) = \partial_{\nu} 
\ln\Phi_G^2$, and the ``local energy'' of the effective 
(``Fixed-Phase'') Hamiltonian $\hat{H}_{FP}$ (see Eq. \ref{fpham} and 
its derivation) is $E_L({\cal R}) = \Phi_G^{-1} \hat{H}_{FP} \Phi_G$ 
with 
\begin{equation}
\hat{H}_{FP} = - D \mbox{\boldmath $\Delta$} + D \sum_{i=1}^{N} 
\left[  (\partial^{\mu} \chi({\cal R}) - \frac{e}{\hbar c} 
a^{\mu}(i))(\partial_{\mu} \chi({\cal R}) - \frac{e}{\hbar c} 
a_{\mu}(i)) \right] + \widehat{V}({\cal R}) \ ,
\label{fixpham}
\end{equation}
where $\chi({\cal R})$ is the phase of the many-body state $\Phi$, 
i.e. $\Phi = |\Phi| \exp[i \chi]$. 
The differential equation satisfied by the distribution function 
$\bar{f}$ is formally equivalent to the one describing Brownian motion 
on a general manifold (including generation and recombination 
processes), and corresponds to a Kramers-Moyal expansion with exactly 
two terms. In fact, we can rewrite the equation above as a 
Fokker-Planck equation for $d N$ continuous stochastic variables 
$\{{\bf r}_i\}_{i=1,\cdots,N}$
\begin{equation}
\partial_{t} \bar{f} = \left \{ \bar{\cal L}_{\rm FP} - 
(\bar{E}_L-E_T) \right \} \bar{f} \ ,
\label{fock}
\end{equation}
where the (time-independent) Fokker-Planck operator $\bar{\cal L}_{\rm 
FP}$ is given by
\begin{equation}
\bar{\cal L}_{\rm FP} \bullet = \sum_{i=1}^{N} \left[ \partial_{\mu} 
\partial_{\nu} \left ( \bar{D}^{\mu \nu}(i) \bullet \right ) - 
\partial_{\mu}  \left ( \bar{D}^{\mu}({\cal R}) \bullet \right ) 
\right ] \ .
\end{equation}
The diffusion matrix (contravariant tensor) $\bar{D}^{\mu \nu}$ and 
drift $\bar{D}^{\mu}$ (which does not transform as a contravariant 
vector) are given by
\begin{eqnarray}
\bar{D}^{\mu \nu} &=& D \ g^{\mu \nu} \\
\bar{D}^{\mu} &=& \bar{D}^{\mu \nu} F_{\nu} + \partial_{\nu} 
\bar{D}^{\mu \nu} - 
\bar{D}^{\mu \nu} \Gamma_{\nu \sigma}^{\sigma} \ ,
\end{eqnarray}
where $\Gamma_{\mu \nu}^{\sigma}$ is the Christoffel symbol of 
the second kind
\begin{eqnarray}
\Gamma_{\mu \nu}^{\sigma} &=&  \frac{1}{2} g^{\sigma \rho} 
\left( \partial_{\mu} g_{\nu \rho} + \partial_{\nu} g_{\mu 
\rho} - \partial_{\rho} g_{\mu \nu} \right) \\
\Gamma_{\nu \sigma}^{\sigma} &=& \frac{1}{2} \partial_{\nu} \ln g \ ,
\end{eqnarray}
and the modified local energy 
\begin{equation}
\bar{E}_L = E_L + \bar{D}^{\mu \nu} \Gamma_{\mu \sigma}^{\sigma} 
F_{\nu} + \partial_{\mu} \left( \bar{D}^{\mu \nu} \Gamma_{\nu 
\sigma}^{\sigma}\right) \ .
\end{equation}
Notice, however, that singularities in the ``quantum corrections'' 
\cite{qc} to the local energy $E_L$ due to the metric, can induce very 
large fluctuations in $\bar{E}_L$. Moreover, the probability density 
$\bar{f}$ does not transform as a scalar function ($\bar{f}({\cal
R},t) \ \bar{\mbox{\boldmath ${\omega}$}} = \bar{f}({\cal R}',t) \ 
\bar{\mbox{\boldmath ${\omega}$}}'$, where the primes represent the 
transformed coordinates and $\bar{\mbox{\boldmath ${\omega}$}} = 
{\rm d}x_1^1 \wedge \cdots {\rm d}x_1^d \wedge \cdots {\rm d}x_{ N}^d$ 
is a volume element in ${\cal M}^{N}$). Therefore, it is more 
convenient to work with a probability density that is a scalar and  
which is defined as
\begin{equation}
f({\cal R},t) = \left[ \prod_{i=1}^N g^{1/2}(i) \right] \ 
\bar{f}({\cal R},t) \ .
\label{ff}
\end{equation}
The differential equation $f$ satisfies is of the form Eq. \ref{fock} 
with bar quantities replaced by unbar ones (e.g. $\bar{\cal L}_{\rm 
FP} \rightarrow {\cal L}_{\rm FP}$). It turns out that $D^{\mu \nu} = 
\bar{D}^{\mu \nu}$ and the drift (which is not a tensor)
\begin{equation}
D^{\mu} = D^{\mu \nu} F_{\nu} + \partial_{\nu} D^{\mu \nu} + 
D^{\mu \nu} \Gamma_{\nu \sigma}^{\sigma} \ .
\label{drift}
\end{equation}
Note that in this case the quantum correction to the local energy 
vanishes. Furthermore, if the metric is diagonal, i.e. $g_{\mu \nu} = 
g^{1/2} \delta_{\mu \nu}$, then the correction to the flat space drift 
also vanishes, i.e. $\partial_{\nu} D^{\mu \nu} + D^{\mu \nu} 
\Gamma_{\nu \sigma}^{\sigma} = 0$, and $D^{\mu} = D g^{-1/2} F^{\mu}$. 
This last remark is quite important, specially for $d=2$ where it is 
{\it always} possible to choose a coordinate system (${\bf r}_i = 
(\xi_i^1,\xi_i^2)$) where the metric tensor is diagonal (conformal 
gauge \cite{polya}), and use the conformal parameterization 
($z_i=\xi_i^1 + i \xi_i^2, \bar{z}_i=\xi_i^1 - i \xi_i^2$) which 
greatly simplifies the resulting expressions (see Section 
\ref{section5}). 

\section{Path Integral Solutions}
\label{section3}

The generalized Fokker-Planck Eq. \ref{fock} describes the time 
evolution of a distribution function $f$ which is completely 
determined by the distribution function at $t=t_0=0$. In this sense it 
describes a continuous stochastic process that is Markovian. Because 
it represents a Markov process, the conditional probability that if 
the system is in $\cal R$ at time $t=0$ it will jump to $\cal R'$ in 
time $t$ (importance-sampled Green's function) $G({\cal R} \rightarrow 
{\cal R}';t)$ contains the complete information about the process, and 
it follows that the probability densities $f({\cal R},t+\tau)$ and 
$f({\cal R},t)$ are connected by
\begin{equation}
f({\cal R}',t+\tau) = \int_{{\cal M}^{N}} \mbox{\boldmath 
$\omega$} \  G({\cal R} \rightarrow {\cal R}';\tau) \ f({\cal R},t) \ ,
\end{equation}
where the Green's function $G({\cal R} \rightarrow {\cal R}';\tau)$ is 
a transition probability for moving particles from ${\cal R}$ to 
${\cal R}'$ in time $\tau$ with the initial value $G({\cal R} 
\rightarrow {\cal R}';0) = \left[ \prod_{i=1}^N g^{-1/2}(i) \right] 
\delta({\cal R}-{\cal R}')$, and is formally given by
\begin{equation}
G({\cal R} \rightarrow {\cal R}';\tau) = \widetilde{\Phi}_G({\cal R}') 
\ \langle {\cal R}' | \exp[ - \tau (\hat{H}_{FP}-E_T)] | {\cal R} 
\rangle \ \widetilde{\Phi}_G^{-1}({\cal R}) \ ,
\end{equation}
with
\begin{equation}
\widetilde{\Phi}_G({\cal R}) = \left[ \prod_{i=1}^N g^{1/2}({\bf r}_i) 
\right] \Phi_G({\cal R}) \ .
\end{equation}
Path integral solutions for $f$ may be derived from the transition 
probability density for small $\tau$. Iteration of the 
Chapman-Kolmogorov equation for $G$ allows one to express the 
evolution of $f({\cal R}',t)$ from the initial distribution $f({\cal 
R},t=0)$ in terms of the short-time Green's function as
\begin{equation}
f({\cal R}',t) = \int_{{\cal M}^{N}} \mbox{\boldmath $\omega$}_{M-1}
\cdots \int_{{\cal M}^{N}} \mbox{\boldmath $\omega$}_0 \ 
G({\cal R}_{M-1} \rightarrow {\cal R}_M;\tau) \cdots G({\cal R}_0 
\rightarrow {\cal R}_1;\tau) \ f({\cal R}_0,0) \ ,
\end{equation}
where $t = M \tau$; we identify ${\cal R}_0 = {\cal R}$ and ${\cal 
R}_{M} = {\cal R}'$. By simple inspection the solution of the 
generalized Fokker-Planck equation $f$ stays positive if it was 
initially positive (i.e., if $f({\cal R},0) > 0$.) 

It is clear then that the functional integral representation of $f$ 
requires knowledge of the infinitesimal evolution operator. It is 
well-known in a similar context \cite{tira} that the integrand of the 
functional integral is not unique, it is discretization dependent 
(compatible with the Markovian property of the paths since ${\cal 
R}(t)$ is only sampled at $t-\tau$ and $t$.) On the other hand, it is 
crucial for numerically simulating those paths to use a discretization 
where the drift velocity and diffusion are evaluated at the {\bf 
prepoint} in the integral equation. Following Feynman \cite{feynman} 
we have determined the functional form of $G({\cal R} \rightarrow 
{\cal R}';\tau)$ to ${\cal O}(\tau^2)$. We are going to present the 
final result and omit the details of the calculation which are just 
simple (although lengthy) manipulations of Taylor series expansions 
and gaussian integration. Thus, to ${\cal O}(\tau^2)$ the short-time 
conditional probability is given by
\begin{eqnarray}
G({\cal R} \rightarrow {\cal R}^\prime; \tau) =
G_{b}({\cal R}\rightarrow{\cal R}^{\prime}; \tau)
\prod_{i=1}^N G^0_i({\cal R} \rightarrow {\cal R}^\prime; \tau)
\label{green}
\end{eqnarray}
where
\begin{eqnarray}
G_{b}({\cal R}\rightarrow{\cal R}^{\prime}; \tau) = 
\exp\left[ -\tau \left( \frac{[ E_L({\cal R}) + E_L({\cal R}^\prime) 
]}{2}-E_T \right) \right] \ ,
\label{green1}
\end{eqnarray}
and
\begin{eqnarray}
G_i^0({\cal R} \rightarrow {\cal R}^\prime; \tau) 
= \left(\frac{1}{4 \pi D \tau} \right)^{d/2}
\exp \left[ - \frac{\left( x'^{\mu}_i - x_i^{\mu} - \tau D^{\mu}({\cal 
R}) \right) g_{\mu \nu}({\bf r}_i) \left( x'^{\nu}_i 
- x_i^{\nu} - \tau D^{\nu}({\cal R}) \right) }{4 D \tau} \right ] \ ,
\label{green2}
\end{eqnarray}
that is, a gaussian distribution with variance matrix $2 D^{\mu \nu}$ 
and mean $x_i^{\mu} + \tau D^{\mu}({\cal R})$. Sample trajectories 
(continuous but nowhere differentiable) are generated by using the 
Langevin equation associated with the process, i.e. $x'^{\mu}_i = 
x_i^{\mu} + \tau D^{\mu}({\cal R}) + \sqrt{\tau} \ \eta$ , where 
$\eta$ is a gaussian random variable with zero mean. Note that 
$D^{\mu}$ and $D^{\mu \nu}$ are evaluated at the {\bf prepoint} in the 
integral equation. 

Therefore, the Wick-rotated path integral for $G({\cal R} \rightarrow 
{\cal R}^\prime; t)$ is
\begin{equation}
G({\cal R} \rightarrow {\cal R}^\prime; t) = \int_{{\cal R}(0)={\cal 
R}}^{{\cal R}(t)={\cal R}'} {\cal D} [\mbox{\boldmath $\omega$}(t)] \ 
\exp [- S \left[{\cal R}(t) \right] ] \ ,
\end{equation}
where the measure ${\cal D} [\mbox{\boldmath $\omega$}(t)] = 
\lim_{M \rightarrow \infty} (4 \pi D \tau)^{-M d/2} \ \mbox{\boldmath 
$\omega$}_1 \cdots \mbox{\boldmath $\omega$}_{M-1}$, and the Euclidean 
action 
\begin{equation}
S \left[{\cal R}(t) \right] = \int_0^t d t' \left \{ \frac{1}{4 D} 
\sum_{i=1}^N     (\dot{x}_i^{\mu} - D^{\mu}[{\cal R}(t')]) \ 
g_{\mu \nu}(i) \ (\dot{x}_i^{\nu} - D^{\nu}[{\cal R}(t')])  
+ E_L[{\cal R}(t')] - E_T \right\} \ .
\end{equation}
The integrand above represents a generalized Onsager-Machlup function 
\cite{onsager}, and the dot is a short-hand for time derivatives. 
In closing this Section, we would like to mention that functional 
integral solutions for $\bar{f}$ can be obtained from previous 
expressions after making the replacement $(D^{\mu},E_L) \rightarrow 
(\bar{D}^{\mu},\bar{E}_L)$.

{\it Interpretation of the quantum corrections.} The general
methodology we have developed so far, can be equally applied to other
situations which do not necessarily involve a curved manifold such as,
for instance, particles moving in a medium with position dependent
diffusion constant. In order to adapt our previous formalism, we need
to understand qualitatively the origin of the quantum corrections to
the short-time propagator obtained above. To this end, we will
illustrate the general idea with the following 1$d$ equation (${\cal
M} = {\rm I\!R}$, $N=1$)
\begin{equation}
\partial_x^2 (D(x) f(x,t)) = \partial_t f(x,t) \ .
\end{equation}
The ``standard'' approach to finding the Green's function for this
problem is simply to solve the equation $\partial_x^2 (D(x) G(x,t)) = 
\partial_t G(x,t)$ subject to the boundary condition $G(x,0) = 
\delta(x)$. We can do this by taking the Fourier transform in $x$:
\begin{equation}
 - \frac{k^2}{2 \pi} \int_{\cal M} d k^\prime \ D(k-k^\prime) \ 
G(k^\prime,t) = \partial_t G(k,t) \ ,
\end{equation}
with the boundary condition $G(k,0) = 1$. This is more complex than 
the usual diffusion equation because we get a convolution of $D(k)$ 
and $G(k,t)$. However, we can find an approximate solution for 
$G(k,t)$, valid for small time, by noting that the boundary condition 
implies that, for small times $t$, $G(k,t) - G(k^\prime,t) \sim {\cal 
O}(t)$ and so we have
\begin{equation}
- \frac{k^2}{2 \pi} \int_{\cal M} dk^\prime \ D(k - k^\prime) \ G(k,t) 
 + {\cal O}(t) = \partial_t G(k,t) 
\end{equation}
or, simply, $- k^2 D(x=0) G(k,t) + {\cal O}(t) = \partial_t G(k,t)$, 
where $D(x=0)$ is $D(x)$ evaluated at the {\bf prepoint}. For small 
times we can ignore the order $t$ term and solve for $G$ with the 
result $G(k,t) = \exp[- t \ k^2 D(0)]$. Taking the inverse Fourier 
transform we finally have
\begin{equation}
G(x,t) = {1\over{\sqrt{4 \pi D(0) t}}} \exp[-x^2/4D(0)t] + {\cal 
O}(t^2) \ ,
\end{equation}
which is just the plain Green's function for a 1$d$ random-walk with 
$D(x)$ evaluated at the {\bf prepoint} and with {\bf no quantum 
corrections}. This is, in fact, the result that the Green's function 
for the Jacobian times $\bar{f}$ has no quantum corrections. 

To make things clear let us look at a different equation with the 
same boundary conditions
\begin{equation}
D(x) \partial_x^2 G(x,t) = \partial_t G(x,t) \ ,
\end{equation}
which characterizes a free Brownian particle in 1$d$. Again, taking 
the Fourier transform we get
\begin{equation}
- \frac{1}{2 \pi} \int_{\cal M} dk^\prime \ {k^\prime}^2 \ D(k - 
k^\prime) \ G(k^\prime,t) = \partial_t G(k,t) \ .
\end{equation}
Making the same approximation as above, replacing $G(k^\prime,t)$
with $G(k,t)$ in the integrand, making an error of ${\cal O}(t)$, and 
noticing that 
\begin{equation}
\frac{1}{2 \pi} \int_{\cal M}  dk^\prime \ (k-k^\prime)^2 \ 
D(k^\prime)  = k^2 D(0) + 2 k i \ \partial_x D(0) - \partial_x^2 D(0)
\end{equation}
we get $G(k,t) = \exp [- t \ (k^2 D(0) + 2 k i \ \partial_x D(0) - 
\partial_x^2 D(0))]$. When we take the inverse Fourier transform the 
terms with derivatives of $D(x)$ at $x=0$ give precisely the quantum 
corrections for this simple case
\begin{equation}
G(x,t) = {1\over{\sqrt{4 \pi D(0) t}}} \exp[-(x- 2 t \partial_x 
D(0))^2/4D(0)t + t \ \partial_x^2 D(0) ] + {\cal O}(t^2) \ .
\end{equation}
This is the result that the Green's function for $\bar{f}$ has quantum 
corrections.

The general case is just as simple, and the following rule
emerges. Given any second order differential equation (first order in
$t$), no matter how many dimensions, with or without curvature, with
or without a position dependent diffusion constant, the rule for
obtaining the short-time Green's function with everything evaluated at
the prepoint is as follows: {\bf Bring all derivatives in each term
all the way to the left of that term}.  Once this is done one can
simply write down the Green's function for a generalized diffusion
process assuming the $D(x)$, $D^{\mu}$, $D^{\mu \nu}$, whatever
position dependent terms they may be, are constant and evaluated at
the prepoint. The quantum corrections are then seen to be simply those
extra terms we get when commuting the derivatives to the left.

{\it Fermion-phase problem: Fixed-Phase method.} 
It is evident that one cannot make a probability density out of a 
complex and/or antisymmetric wave function. This is the reason why we 
decided to write down Fokker-Planck equations for the distribution $f$ 
(or $\bar{f}$) and not $\tilde{f}$. Nevertheless, the phase factor 
associated with the original complex distribution must show up in the 
evaluation of the expectation values. It is well-known that this 
causes the variance of the computed results to increase exponentially 
with increasing number of degrees of freedom. This problem is known 
as the fermion-phase \cite{Note1} catastrophe, and in this Section we 
will review a method \cite{fp} to obtain stable, albeit approximate, 
path-integral solutions whose stochastic determination has a 
polynomial, instead of exponential, complexity. The generalization of 
the ideas presented below to bosons (with complex-valued states) or 
anyons in general is straightforward \cite{Wilczek}.  

In order to avoid cumbersome notation which would obscure the main
ideas, here we will consider a simplified version of our original
Hamiltonian
\begin{equation}
\widehat{\rm I\!H}_{\cal R} = \frac{{\bf \Pi} \cdot {\bf \Pi}}{2 m^*} 
+ \widehat{V}({\cal R}) \ ,
\end{equation}
where, for simplicity, we introduced the vector notation, ${\bf \Pi} = 
\left( {\bf \Pi}(1), \cdots, {\bf \Pi}(N) \right)$, and in this 
subsection the same convention will be used for other bold quantities. 
The microscopic equations governing the imaginary-time evolution of 
our interacting system can be found from a variational principle of 
the form $\delta S[{\cal R}(t)] = 0$, where the Euclidean action is 
given by
\begin{equation}
S[{\cal R}(t)] = \int_{0}^{t} d t' \int_{{\cal M}^{N}} \mbox{\boldmath 
$\omega$} \left\{ \frac{1}{2 m^*} \left[ {\bf \Pi} \Phi \right]^* 
\cdot \left[ {\bf \Pi} \Phi \right] + \widehat{V} \ \Phi^* \Phi + 
\Phi^* \partial_{t'} \Phi \right\} \ .
\end{equation}

Finding the states $\Phi$ and $\Phi^*$ which minimize $S[{\cal R}(t)]$ 
is equivalent to solving the Schr\"odinger equation and its complex 
conjugate. Equivalently, we can consider as independent real fields 
the phase and modulus of the wave function, i.e. $|\Phi|$ and $\chi$ 
such that $\Phi = |\Phi| \exp[i \chi]$, and perform independent 
functional variations on $S[{\cal R}(t)]$. The resulting 
Euler-Lagrange equations are:
\begin{equation}
 \left\{  \begin{array}{rclcl}
- \partial_t |\Phi| &=& 
\mbox{ $\Re e \left\{ \exp \left[-i \chi \right] 
\widehat{\rm I\!H}_{\cal R}  \Phi \right\}$} & = &
 \hat{H}_{FP} \; |\Phi|\\
         &     & \\
- (\partial_t \chi) \ |\Phi|^2 &=& 
\mbox{ $\Im m \left\{ \exp \left[-i \chi \right] 
\widehat{\rm I\!H}_{\cal R}  \Phi \right\}$} |\Phi| & = & - 
\frac{\hbar}{2} \ \mbox{\boldmath ${\partial}_{\mu}$} \cdot 
{\bf J}^{\mu}
\label{onep}
          \end{array} 
\right. ,
\end{equation}
where $\Re e$ and $\Im m$ stand for the real and imaginary parts of 
the expressions in brackets, respectively,  
\begin{equation}
\hat{H}_{FP} = \frac{{\bf p} \cdot {\bf p}}{2 m^*} + 
\widetilde{V}({\cal R}) , \; \widetilde{V} = \widehat{V} + D \ 
(\mbox{\boldmath $\partial$}^{\mu} \chi({\cal R}) - \frac{e}{\hbar c} 
\mbox{\boldmath $a$}^{\mu}) \cdot (\mbox{\boldmath $\partial$}_{\mu} 
\chi({\cal R}) - \frac{e}{\hbar c} \mbox{\boldmath $a$}_{\mu}) \ ,
\label{fpham}
\end{equation}
and
\begin{equation}  
{\bf J}^{\mu} = \frac{1}{2 m^*} \left( \Phi^* 
\left[ {\bf \Pi}^{\mu} \Phi \right] + \left[ {\bf \Pi}^{\mu} \Phi 
\right]^* \Phi \right) = \frac{|\Phi|^2}{m^*} \ \left(\hbar 
\mbox{\boldmath $\partial$}_{\mu} \chi({\cal R}) - \frac{e}{c} 
\mbox{\boldmath $a$}_{\mu} \right)
\end{equation}
is the probability current. The singularities of $\chi({\cal R})$ 
occur at the zeros of $\Phi$, which generically have codimension 
two. 

So far we have simply mapped the original fermion problem into a 
bosonic one for $|\Phi|$ but still coupled to its phase fluctuations. 
Alternatively, one can regard this as a gauge transformation of the 
original fermion problem, whose effect is to add a non-local gauge 
field potential, $\mbox{\boldmath $\partial$}^{\mu}\chi$, giving 
rise to a {\it fictitious} magnetic field. Notice that it is this 
gauge field that contains information on particle statistics. 
Moreover, although the geometry of $\chi({\cal R})$ can be 
altered by a gauge transformation, the singularities remain 
invariant. 

The Fixed-Phase (FP) method \cite{fp} consists in making a choice for
the phase, $\chi_T$, and solving the bosonic problem for $|\Phi|$ {\it
exactly} using stochastic methods. The method is {\it stable} and has
the property of providing a variational upper bound to the exact ground
energy $E_0$, $E_{FP} \geq E_0$ (the equality holds when $\chi_T$ is
the exact ground state phase), and for a given $\chi_T$ the lowest
energy consistent with this phase. The trial phases $\chi_T$ should
conserve the symmetries of $\widehat{\rm I\!H}_{\cal R}$ and particle
statistics (for time-reversal invariant systems there is a way for
systematically improving a given mean-field phase using projection
techniques \cite{vortex}). Notice that the FP method projects out the
lowest energy state of a given symmetry. Therefore, the method allows
one to compute also excitations which are ``ground states'' of a
particular symmetry. For ground state properties of real symmetric
Hamiltonian operators the FP approach reduces \cite{fp} to the
Fixed-Node method \cite{reynolds}. 

\section{Computational Implementation}
\label{section4}

In this Section we present an algorithm for computing the ground state
properties of quantum many-body systems defined on a curved manifold
with general metric $g^{\mu \nu}$. As mentioned in the Introduction,
this can be accomplished by performing all multidimensional integrals
using Monte Carlo techniques. In this way, ground state expectation
values are obtained by averaging over a large number of particle
configurations generated according to a certain limiting probability
distribution $p({\cal R}, t\rightarrow \infty)$. There is some freedom
in the choice of this distribution $p({\cal R}, t)$, however, to
reduce statistical fluctuations in the observables to be computed it
is more efficient to use the so-called {\it importance-sampled}
distribution $\bar{f}({\cal R}, t)$, which is the product of the
absolute value of the solution of the time dependent Schr\"odinger
equation $\Phi({\cal R},t)$ and some positive function $\Phi_G({\cal
R})$ that is the best available approximation to the modulus of the
ground state eigenfunction. In a curved manifold, on the other hand,
it is more convenient to work with the {\it modified
importance-sampled} distribution $f({\cal R}, t)$, which is defined as
a product of the conventional importance-sampled distribution
$\bar{f}({\cal R}, t)$ and the metric (see Eq. \ref{ff}).

The propagation of particle configurations in time $\tau$ is 
determined by the 
conditional probability (Green's function) $G({\cal R}\rightarrow{\cal 
R}^{\prime}; \tau)$, whose separation into a diffusion (plus drift) 
and branching parts (see Eq. \ref{green}) makes it very simple to 
simulate numerically. The gaussian term represents propagation 
according to the equation $x'^{\mu}_i = x_i^{\mu} + \tau D^{\mu}({\cal 
R}) + \sqrt{\tau} \ \eta$ , where $\eta$ is a gaussian random 
variable with zero mean. The effect of the term $\tau D^{\mu}({\cal 
R})$ is to superimpose a drift velocity on the random diffusion 
process so that particle configurations are directed towards regions 
of configuration space where $\Phi_G({\cal R})$ is large. The 
branching term $G_b({\cal R}\rightarrow{\cal R}^{\prime}; \tau)$ in 
Eq. \ref{green1}, determines the creation and annihilation of 
configurations (walkers) at the point ${\cal R}^{\prime}$ after a 
move. If the size of the ensemble of walkers at any time $t$ is 
defined as 
\begin{equation}
{\cal P}(t)=\int_{{\cal M}^{N}} \bar{\mbox{\boldmath $\omega$}} \ 
f({\cal R},t) 
\end{equation}
then, its rate of change is given by
\begin{equation}
\partial_t {\cal P}(t)= - \int_{{\cal M}^{N}} 
\bar{\mbox{\boldmath $\omega$}} \ \left[ E_L({\cal R}) - E_T 
\right] f({\cal R},t) \ .
\end{equation}
Therefore, if the local energy $E_L({\cal R})$ is a smooth function 
of ${\cal R}$, and the trial energy $E_T$ is suitably adjusted, the 
size of the ensemble of walkers will remain approximately constant 
as the configurations propagate. In particular, if the local energy 
is constant and equal to $E_T$ then the fluctuations in the 
ensemble size will vanish.  
To ease notation, in the rest of the paper we will only consider the 
standard situation $\Phi_T=\Phi_G$. In such a case, ground state 
expectation values of a generic observable $\widehat{\cal O}$ will be 
computed as 
\begin{equation}
\lim_{t \rightarrow \infty} \frac{\langle \Phi_G | \ \widehat{\cal O} 
\ \hat{\cal U}(t) \ \Phi_G \rangle}{\langle 
\Phi_G | \ \hat{\cal U}(t) \ \Phi_G \rangle} = 
\langle \Phi_G^{-1} \widehat{\cal O} \Phi_G \rangle_{f(t 
\rightarrow \infty)}= \int_{{\cal M}^{N}} 
\bar{\mbox{\boldmath $\omega$}} \ \frac{f({\cal R},t \rightarrow 
\infty)}{{\cal P}(t \rightarrow \infty)} 
\ [\Phi_G^{-1} \widehat{\cal O} \Phi_G]({\cal R})\ .
\end{equation}

In the following we present a step by step computational algorithm for
implementing the stochastic approach discussed above. Most parts of the
algorithm follow closely the standard DMC method, described for
instance in \cite{reynolds}, but we believe that it is useful to
present these steps in detail here because a number of straightforward
but subtle modifications due to the space curvature are involved.  To
be more specific (without loosing the general features), let us present
the algorithm as it is applied to fermionic systems within the FP
approach.

{\it Algorithm.} 

$\S$ 1. Construct a guiding wave function $\Phi_G$, which is the best
available approximation to the exact ground state (or lowest energy
state of a given symmetry). In principle any choice of $\Phi_G$ which
has finite overlap with the exact state is acceptable, but the more
accurate $\Phi_G$ is the faster the convergence to the stationary
solution will be, and the lower the statistical fluctuations will be
as well. Recall that zero variance is obtained when the guiding wave
function is equal to the desired ground state (and the ground state is
bosonic).

$\S$ 2. Given the guiding function compute the quantum drift velocity 
$F_{\nu}= \partial_{\nu} \ln \Phi_{G}^{2}$ and the local energy 
$E_L=\Phi_{G}^{-1}{\hat H}_{FP} \Phi_{G}$, where ${\hat H}_{FP}$ is 
the FP Hamiltonian defined in Eq. \ref{fixpham}. $F_{\nu}$ 
is used to construct the drift vector $D^{\nu}$ according to Eq. 
\ref{drift}, and the drift and local energy are used in evaluating 
the short-time Green's function according to Eqs. \ref{green}, 
\ref{green1}, and \ref{green2}. When a simple guiding function can 
be constructed, the expressions for the drift and the local energy can 
be evaluated analytically, but in most cases this must be done 
numerically.

$\S$ 3. A set of $N_w$ initial configurations or walkers $\{{\cal R}_j
(t=0)\}$ ($j=1,\cdots,N_w$) is created, such that particles in each 
walker are distributed according to the {\it modified 
importance-sampled} distribution $f({\cal R},t=0) = \left[ 
\prod_{i=1}^N g^{1/2}(i) \right] |\Phi_G|^2$ (which is equivalent to 
using the Variational Monte Carlo (VMC) technique).

$\S$ 4. Each walker ${\cal R}_j$ is diffused for a time $\tau$ 
according to the gaussian part of the propagator $\prod_{i=1}^N 
G^0_i({\cal R}_j \rightarrow {\cal R}_j^\prime; \tau)$. This can be 
accomplished by moving each particle coordinate $x_{i}^{\mu}$ 
according to 
\begin{equation}
x_{i}^{\prime \mu} = x_{i}^{\mu} + \tau D^{\mu}({\cal R}_j)+ 
\sqrt{\tau} \ \eta \ ,
\end{equation}
where $\eta$ is a Gaussian random variable with a mean of zero and a 
variance of $2D^{\mu \nu}=2D g^{\mu \nu}$. 

$\S$ 5. The move from ${\cal R}_j$ to ${\cal R}^{'}_j$ is then 
accepted with a probability
\begin{equation}
A({\cal R}_j \rightarrow {\cal R}_j^{\prime}; \tau) \equiv 
\min(1,W({\cal R}_j, {\cal R}_j^{\prime}; \tau)) \ ,
\end{equation}
where 
\begin{equation}
W({\cal R}, {\cal R}^{\prime}; \tau) \equiv  \left[ \prod_{i=1}^N
\frac{g({\bf r}^{'}_i)}{g({\bf r}_i)} \right] \left|
\frac{\Phi_G({\cal R}^{'})}{\Phi_G({\cal R})} \right|^2
\frac{G({\cal R}^{\prime} \rightarrow {\cal R}; \tau)}{G({\cal R} 
\rightarrow {\cal R}^{\prime}; \tau)} \ .
\end{equation}
This step ensures detailed balance in the Monte Carlo procedure. A 
typical acceptance ratio is in excess of 99\%. Notice that if 
$G({\cal R} \rightarrow {\cal R}^{\prime}; \tau)$ is the exact 
Green's function, and not its short-time approximation, then $W$ is 
unity and this step is not necessary. This is because the Green's 
function of an Hermitian operator is symmetric, i.e. 
$G({\cal R} \rightarrow {\cal R}^{\prime}; \tau) = G({\cal R}^{\prime} 
\rightarrow {\cal R}; \tau)$, but this is not the case for any 
importance-sampled distribution function equation.  

$\S$ 6. After all the particles of a given walker have been diffused 
from the initial position ${\cal R}_j$ to the position 
${\cal R}_j^{\prime}$ and the move is accepted, then the values of 
the drift $D^{\mu}$, the local energy $E_L$, and $\Phi_G$ are updated.
We could have equally well move all particles at once in step 4 
before step 5, however, the acceptance probability for a given 
time step is reduced considerably. Depending upon the particular 
problem one can adopt one of the two strategies: single or 
multiparticle moves. 

$\S$ 7. The multiplicity (weight) $\mathsf{M}$ of a given walker, is 
computed from the branching part of the Green's function Eq. 
\ref{green1}
\begin{equation}
{\mathsf{M}}= G_b({\cal R} \rightarrow {\cal R}^{\prime}; \tau_a) \ ,
\end{equation}
where $\tau_{a} \leq \tau$ because some of the moves can be rejected. 
If the mean-squared distance the particles would diffuse in the 
absence of the rejection step is $\langle r^2 \rangle_{tot}$, and the 
actual mean-squared distance is $\langle r^2 \rangle_{a}$ then the 
actual time used in a branching step $\tau_{a}= \tau \langle r^2 
\rangle_{a}/\langle r^2 \rangle_{tot}$. In the case of multiparticle 
moves, the effective time step must be calculated separately from a 
computation of the accepted to attempted moves. Since $\mathsf{M}$ is, 
in general, not an integer one can use instead the integer 
\[\hat{\mathsf{M}}={\rm int}(\mathsf{M} + \xi) \ ,\] 
where $\xi$ is a random number uniform in the range $[0,1]$. In this 
case the average density of walkers is conserved and $\langle 
\hat{\mathsf{M}} \rangle=\mathsf{M}$. If $\hat{\mathsf{M}}=0$ then the 
walker is deleted from the ensemble; otherwise $\hat{\mathsf{M}}-1$ 
copies of the configuration are made and added to the ensemble. 
Note that fixing $\hat{\mathsf{M}}=1$ is equivalent to eliminate 
branching and, consequently, to perform a VMC 
calculation with limiting distribution $f({\cal R},t \rightarrow 
\infty)= \left[ \prod_{i=1}^N g^{1/2}(i) \right] |\Phi_G|^2$. 

$\S$ 8. After diffusing and branching all walkers in time $\tau$ the 
mean value of the local energy over all walkers is computed from the 
obtained distribution. One has to start averaging over these values, 
after some target (equilibration) time, when the configurations are 
sampled according to the limiting stationary distribution $f({\cal R}, 
t\rightarrow \infty)$. The target time depends upon the particular 
problem and how close $\Phi_G$ is to the exact state.   

$\S$ 9. Using the average value of the local energy over the whole 
ensemble of walkers $\langle E_L({\cal R}) \rangle_{f(t \rightarrow 
\infty)}$, the trial energy $E_T$ is updated according to $E_T = 
(E_T+E_L)/2$. Mixing this estimate with an old value of the trial 
energy, allows one to improve the convergence.     

$\S$ 10. After computing the average energy over a sufficiently long 
time ($\tau N_m$, where $N_m$ is the number of moves per 
configuration), its value is stored and the first block is completed. 
$N_m$ should be large enough for there to be little statistical 
correlation between energy subaverages obtained in different blocks. 
A new ensemble of $N_w$ walkers is generated by randomly copying or 
deleting configurations, and steps 4 to 9 are repeated completing 
another block. One has to do as many blocks as it takes in order to 
reach the desired statistical accuracy. Notice, however, that because 
the propagator is only accurate to ${\cal O}(\tau^2)$, the 
distribution $f({\cal R},t \rightarrow \infty)$ and the resulting 
estimates will have a time step bias. The way to eliminate this bias 
is by extrapolating all computed expectation values to $\tau 
\rightarrow 0$. 

\begin{figure}
\centerline{\psfig{figure=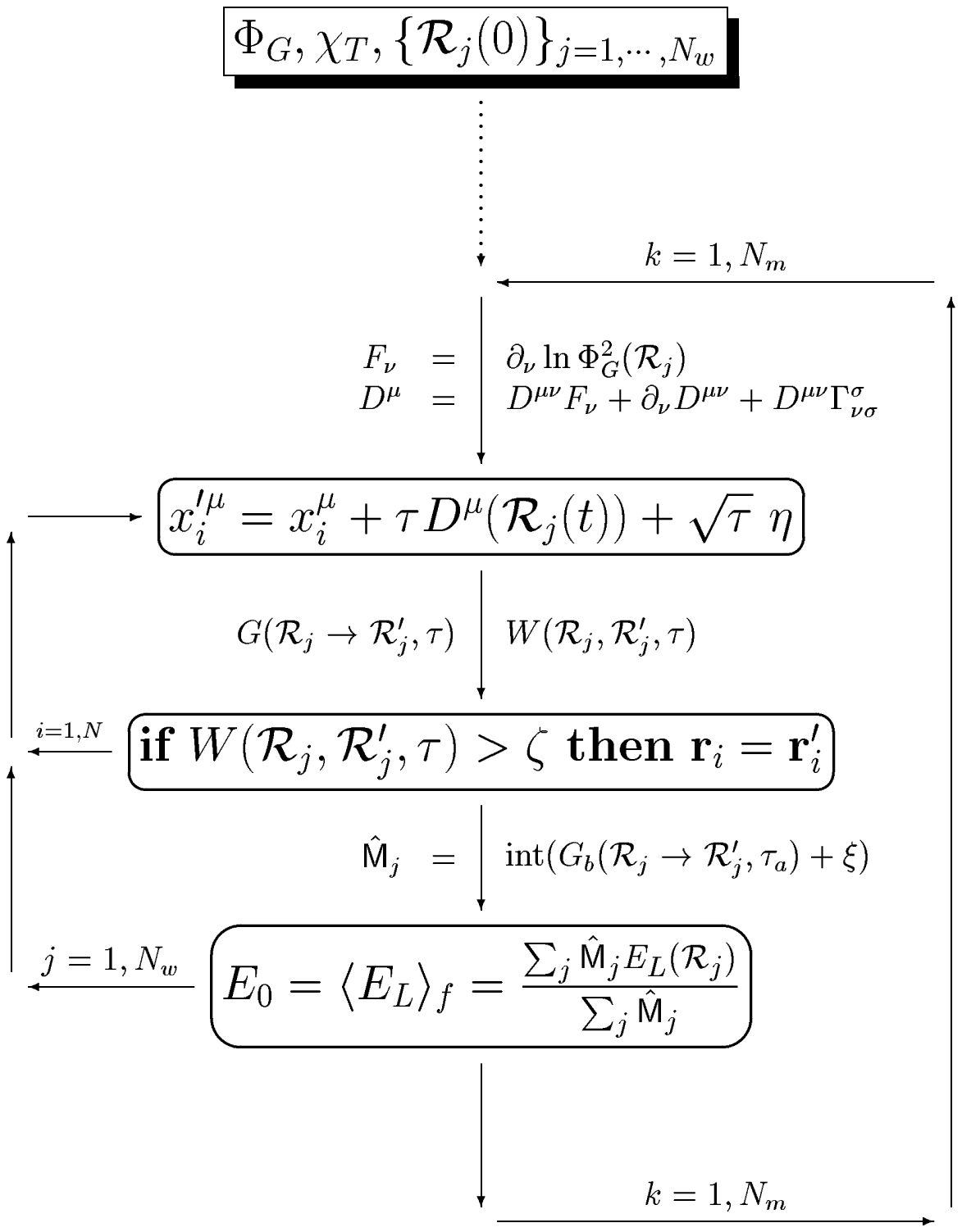,height=8in,angle=0}}
\caption{A schematic of the Fixed-Phase method for curved manifolds 
with general metric $g^{\mu \nu}$. See the text for notation.} 
\label{fig0}
\end{figure}

Figure \ref{fig0} shows a schematic flow diagram of the algorithm 
presented in this Section. 

\section{Example: Electron-Monopole in S$^2$}
\label{section5}

As an example application of the method we have developed in the
previous Sections we consider the problem of a single particle of
charge $e$, mass $m^*$ and vector position ${\bf r} = (x^1,x^2,x^3)$
in ${\rm I\!R}^3$ confined to the surface of a sphere of radius $R$
centered at the origin (${\cal M}$ = S$^2$, $N=1$) moving in the
presence of the vector potential of a Dirac monopole at the origin.
This problem can be solved in closed form and so constitutes an ideal
model system for testing the accuracy of the stochastic solutions we
can derive using the formalism developed in previous
Sections. Moreover, the case of $N$ interacting electrons confined to
S$^2$ in the presence of a monopole field serves as the basic model
which captures the essential physics of the quantum Hall effect
\cite{Haldane}.

\begin{figure}
\centerline{\psfig{figure=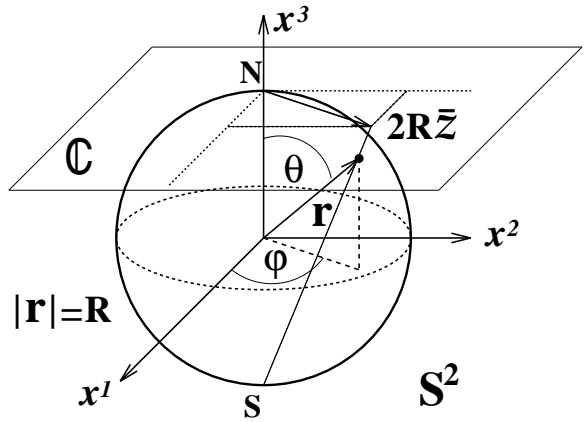,height=3in,angle=270}}
\vspace{1in}
\caption{Spherical and stereographic projection coordinates. $R$ is
the radius of the two-sphere S$^2$. Notice that points on the sphere 
are projected into the complex plane ($z \in \bbbc$) from the southern 
pole.} 
\label{fig1}
\end{figure}

The Pauli Hamiltonian for a spinless particle in S$^2$ is
\begin{equation}
\widehat{\rm I\!H}_{\cal R} = \frac{\left | \hat{\bf r} \wedge (-i 
\hbar \nabla -(e/c) \ {\bf A}) \right|^2}{2 m^*} \ ,
\end{equation}
where $\hat{\bf r} = {\bf r}/R$, and ${\bf A}$ is the monopole vector 
potential ($\nabla \wedge {\bf A} = B \ \hat{\bf r}$, $B$ being the 
strength of the radial field.) Therefore, the total number of flux 
quanta $2{\cal S}$ piercing the surface of the sphere is given by 
$2{\cal S} = 4 \pi R^2 B / \phi_0$, where $\phi_0=hc/|e|$ is the 
elementary flux quantum. Following Wu and Yang \cite{Wu} we can 
construct angular momentum operators ${\bf L} = {\bf r} \wedge (-i 
\hbar \nabla -(e/c) \ {\bf A}) + \hbar {\cal S} \ \hat{\bf r}$ in
terms of which the Hamiltonian reads
\begin{equation}
\widehat{\rm I\!H}_{\cal R} = \frac{|{\bf L}|^2 - \hbar^2 {\cal 
S}^2}{2 m^* R^2} \ .
\label{mono}
\end{equation}
If we choose a gauge where the vector potential is ${\bf A}= - B R \ 
\cot \theta \ \hat{\varphi}$, then the Hamiltonian, Eq.~\ref{mono}, 
can be written as    
\begin{equation}
\widehat{\rm I\!H}_{\cal R} = \frac{D}{R^2} \left[ - 
\partial^{2}_{\theta} - \frac{1}{\sin^2\theta} \ 
\partial^{2}_{\varphi} - \cot\theta  \ \partial_{\theta}
+ 2 i {\cal S} \ \frac{\cot\theta}{\sin\theta} \ \partial_{\varphi} +
{\cal S}^2 \ \cot^2\theta \right] \ ,
\end{equation}
in terms of the usual spherical angles $\theta$ and $\varphi$ (
$0 \leq \theta \leq \pi$, $0 \leq \varphi < 2 \pi$, see 
Fig.~\ref{fig1}.) The eigenstates of this Hamiltonian are monopole 
harmonics (normalized to 1)\cite{Wu}
\begin{eqnarray}
Y_{{\cal S},n,m} &=& {\cal N}_{{\cal S}nm} \ (-1)^{{\cal S}+n-m} \ 
\exp[-i{\cal S}\varphi] \ u^{{\cal S}+m} \ v^{{\cal S}-m} 
\ {\cal F}(|u|,|v|) \ , \nonumber \\
{\cal F}(|u|,|v|) &=& \sum_{k=0}^{n} (-1)^k {n\choose k}{ 
{2{\cal S}+n} \choose {{\cal S}+n-m-k}} (v \bar{v})^{n-k} (u  
\bar{u})^{k} \ , \nonumber \\
{\cal N}_{{\cal S}nm} &=& \left(\frac{2{\cal S}+2n+1}{4 \pi} \ 
\frac{({\cal S}+n-m)! \ ({\cal S}+n+m)!}{n! \ (2{\cal S}+n)!}
\right)^{1/2} \ ,
\end{eqnarray}
where $u=\cos(\theta/2) \exp[i\varphi/2]$ and $v=\sin(\theta/2)
\exp[-i\varphi/2]$ are spinor coordinates, $n$ is the Landau level 
quantum number, and $m = -{\cal S}-n,-{\cal S}-n+1, \cdots, {\cal 
S}+n$ is the ($L_{x^3}$) angular momentum quantum number which labels 
degenerate states within the $n^{\rm th}$ level. In the sum 
above the binomial coefficient ${\alpha \choose \beta}$ vanish when 
$\beta > \alpha$ or $\beta < 0$. For a given ${\cal S}$ and $m$, the 
ground state ($n=0$) and first excited state ($n=1$) are 
\begin{eqnarray}
\psi_{gs} &\propto& u^{S+m}v^{S-m} \ , \\
\psi_{es} &\propto& \left[2({\cal S}+1) v \bar{v} - ({\cal S}-m+1) 
\right] \ \psi_{gs} \ ,
\end{eqnarray} 
respectively. The energy of a state with Landau level quantum number 
$n$ is given by
\begin{equation}
E_n= \left( 2n+1+\frac{n(n+1)}{\cal S} \right) \ \frac{\hbar 
\omega_c}{2} \ , 
\end{equation}
where $\omega_c$ is the cyclotron frequency ($\omega_c=|e|B/m^*c$).

Let us now reformulate the electron-monopole problem in a way 
consistent with the notations introduced in the previous Sections.
In this way one can compare the exact result to the numerical one 
obtained with the algorithm developed in this paper, thus testing the 
numerical technique. First, instead of the spherical angles $\theta$ 
and $\phi$ we introduce new coordinates $z$ and $\bar{z}$, where $z= 
\tan(\theta/2) \exp[-i\varphi]$, and $\bar{z}$ is its complex 
conjugate. Geometrically, this transformation can be viewed as a 
stereographic projection of the sphere onto the plane, as illustrated 
in Fig.~\ref{fig1}. The Hamiltonian can be rewritten as 
 \begin{equation}
\widehat{\rm I\!H}_{\cal R} = \frac{i}{m^*} \ g^{-1/4} ( p_z 
- \bar{\cal A}(z) ) \left( p_{\bar{z}} - {\cal A}(z) \right) 
g^{1/4} + \frac{D {\cal S}}{R^2} \ ,
\end{equation}
in terms of the (non-hermitian) canonical momenta $p_z$ and 
$p_{\bar{z}}$, and 
\begin{equation}
{\cal A}(z) = -i \frac{\hbar {\cal S}}{2} z \ \left( 
\frac{1-|z|^2}{|z|^2 \ (1+|z|^2)} \right) \ ,
\end{equation}
with metric tensor
  \begin{equation}
g^{\mu \nu}(z,\bar{z}) = \pmatrix{0&\frac{(1+ z \bar{z})^2}{2 R^2}\cr
                    \frac{(1+z \bar{z})^2}{2 R^2}&0\cr} \ .
\end{equation}
Naturally, the particles moving in the projected plane are in a space 
with curvature, corresponding to that of the sphere. Notice that the 
metric tensor is diagonal when written in terms of ($\xi^1,\xi^2$), 
such that $z=\xi^1+ i \ \xi^2$, i.e. $g^{\mu \nu}(\xi^1,\xi^2)= 
\frac{(1+|z|^2)^2}{4 R^2} \ \delta^{\mu \nu}$ (i.e, it corresponds 
to the conformal gauge). Then, the drift is simply $D_{\mu}=D \ 
F_{\mu}$. 

The stochastic method developed above allows one to obtain the {\rm
exact} energy eigenvalues of the electron-monopole problem in S$^2$
iff we know the exact phase of the eigenfunctions. In other words, if
the trial state is chosen such that it has the exact ground state
phase, then independently of its modulus our stochastic approach will
lead to the exact ground state energy. Similarly, if the trial state
has a phase corresponding to an excited state eigenfunction then we
will obtain the exact excited state energy eigenvalue. In the next two
subsections we construct simple trial states for the ground and first
excited states of the one particle problem. Their modulus are then
used as guiding functions $\Phi_G$. Using these trial states we will
apply our technique and illustrate the main ideas of our method.

{\it Ground State} $(n=0)$: The $2{\cal S}+1$ degenerate ground states 
of the electron-monopole system are labeled by their $L_{x^3}= \hbar
[\bar{z}  \partial_{\bar{z}} - z \partial_z]$ angular momentum quantum
numbers $m = -{\cal S}, \cdots ,{\cal S}$. Here we consider the
$m={\cal S}$ ground state for which the exact (unnormalized) wave
function can be written
\begin{eqnarray}
\psi_{gs} = \left( \frac{|z|}{z(1+|z|^2)} \right)^{\cal S} \equiv
|\psi_{gs}| e^{i\varphi_{gs}} \ .
\end{eqnarray}
To illustrate our method we imagine that we do not know this exact
ground state $\psi_{gs}$ but instead we have constructed the following
two trial states
\begin{equation}
\psi_{T1}=\left( \frac{|z|}{z (1 + |z|^2)} \right)^{\cal S}
\frac{1}{1+\lambda |z|^2} \equiv |\psi_{T1}| e^{i\varphi_{T1}} \ ,
\end{equation}
and
\begin{equation}
\psi_{T2}=\left( \frac{|z|}{z (1 + |z|^2)} \right)^{\cal S} \left(
\frac{|z|}{z} \right)^{\alpha} \equiv |\psi_{T2}| e^{i\varphi_{T2}} \ ,
\end{equation}  
where $\lambda$ and $\alpha$ are real valued constants.

The trial states $\psi_{T1}$ and $\psi_{T2}$ have been constructed so
that for $\lambda=0$ and $\alpha = 0$ they are both equal to the exact
ground state, $\psi_{gs}$.  For $\lambda \ne 0$, the modulus of
$\psi_{T1}$ is no longer equal to that of the exact ground state, but
the phase of the wave function is {\it exact}, i.e.
\begin{eqnarray}
|\psi_{T1}| \ne |\psi_{gs}| \ ,~~~~~~~~~~~ \varphi_{T1} =
\varphi_{gs} \ .
\end{eqnarray}
In contrast, for $\alpha \ne 0$, the modulus of $\psi_{T2}$ is exact,
but the phase is approximate,
\begin{eqnarray}
|\psi_{T2}| = |\psi_{gs}| \ ,~~~~~~~~~~~~\varphi_{T2} \ne
\varphi_{gs} \ .
\end{eqnarray}
It follows that if $\psi_{T1}$ is used as a trial state in a FP DMC
simulation the resulting energy will be the {\it exact} ground state
energy $E_0 = \hbar \omega_c/2$, while if $\psi_{T2}$ is used as a
trial state the simulation will not lead to the exact ground state
energy, but instead will provide a variational upper bound.

As has already been emphasized the trial state used in a FP DMC
simulation should be constructed to be the {\it best available}
approximation to the exact eigenstate, since the quality of the trial
state can greatly influence the speed of convergence and the
statistical accuracy of the result of the simulation. This can be
clearly illustrated by considering the trial state $\psi_{T2}$. Since
the modulus of $\psi_{T2}$ is exact the drift velocity $F$ which
depends only on the modulus of the trial state will correspond to the
exact drift velocity and in the absence of the branching term will
lead to the exact density distribution. It is straightforward to show
that
\begin{eqnarray}
F_1 = -\frac{4{\cal S} \ \xi^1}{1+|z|^2} \ , ~~~ 
F_2 = -\frac{4{\cal S} \ \xi^2}{1+|z|^2} \ ,
\end{eqnarray}  
where $F_{\mu} = \partial_{\xi^{\mu}} \ln |\psi_{T2}|^2$, indicating
that walkers are guided away from the regions where the wave function
is small and, in this way, the particle tends to spend most of the time
near the top of the sphere ($\theta=0$). A potential problem appears
when we consider the local energy,
\begin{eqnarray}
E_L= |\psi_{T2}|^{-1} \hat{H}_{FP} |\psi_{T2}| = \frac{\hbar
\omega_c}{2} \left [ 1 + \frac{(1+|z|^2)^2}{\cal S} \left (
\frac{\alpha {\cal S}}{1+|z|^2} + \frac{\alpha^2}{4 |z|^2} \right) \right] 
\end{eqnarray}
which, of course, is not exact due to the approximate phase of the
trial state.  In particular, $E_L$ {\it diverges} as $|z| \rightarrow
0$. As we have just shown, the drift will tend to push the particle
towards $z=0$, leading to large fluctuations in the local energy. This
in turn can lead to huge fluctuations in the population size (number
of walkers), since the size of the population depends exponentially on
the local energy. Thus, in this particular example, one has to take
small values for $\alpha$ ($\alpha << 1$) in order to assure fast
convergence and good statistical accuracy.

\begin{figure}
\centerline{\psfig{figure=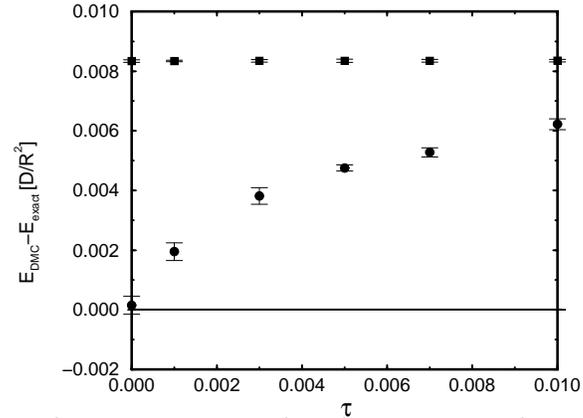,height=3in,angle=270}}
\caption{The difference between computed and exact ground state
energies for the trial state with the exact phase $\psi_{T1}$
(circles) and the trial state with an approximate phase $\psi_{T2}$
(squares) for various values of time step $\tau$. The $\tau = 0$
extrapolated results are also displayed. Using a trial state with the
exact phase in the FP DMC simulations allows one to solve the
problem exactly, while using a trial state with an approximate phase
allows one to obtain a variational upper bound for the exact
solution.}
\label{fig2}
\end{figure}

Figure \ref{fig2} shows the results of FP DMC simulations, using the
algorithm developed in this paper, for the difference between computed
and exact ground state energies for trial state $\psi_{T1}$ (circles)
and trial state $\psi_{T2}$ (squares) using different values of the
time step $\tau$.  The $\tau \rightarrow 0$ extrapolated values are
also shown. For trial state $\psi_{T1}$ we used $\lambda=1$, the
number of walkers was chosen to be $N_w=300$, the number of Monte
Carlo steps per walker was $2 \times 10^7$, and the acceptance rate
was between $97 \%$ and $99.5 \%$. For trial state $\psi_{T2}$ we used
$\alpha=0.001$, the number of walkers was $N_w=100$, the total number
of Monte Carlo steps per walker was $10^6$, and the acceptance rate
was the same as for $\psi_{T1}$.  Since the parameter $\alpha$ in
$\psi_{T2}$ was chosen so that $\alpha \ll 1$, the trial state and the
corresponding Green's function were nearly exact and we were able to
reach reasonable statistical accuracy with a relatively small number
of Monte Carlo steps.  As expected we find that when trial state
$\psi_{T1}$ is used the extrapolated energy agrees within statistical
accuracy with the exact result, but when we use trial state
$\psi_{T2}$, for which the phase is not exact, we obtain a variational
upper bound for the exact ground state energy.

\begin{figure}
\centerline{\psfig{figure=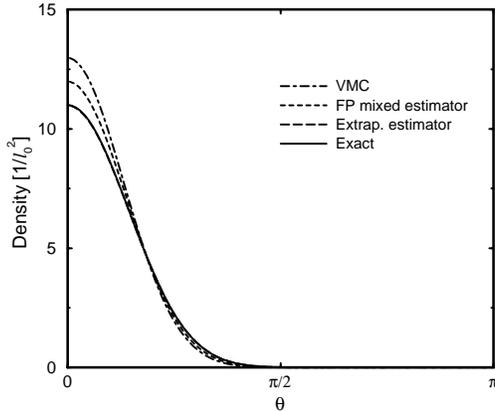,height=3in,angle=270}}
\caption{Ground state density for the exact ground state
$\psi_{T1}(\lambda=0)$ (Exact), the trial state $\psi_{T1}(\lambda=1)$
(VMC), the density obtained in FP diffusion Monte Carlo with trial
state $\psi_{T1}(\lambda=1)$ at time step $\tau=0.001$ (FP mixed
estimator), and the extrapolated density defined as ratio of the square
of FP density to the variational density. The diffusion Monte Carlo
density (FP mixed estimator) improves on the variational result but
still differs from the exact one. The extrapolated estimator for the
density constructed by combining both, the FP mixed estimator and the
variational density makes it possible to improve on FP density and is
very close to the exact result. The density is normalized in such a way
that its integral over the surface of the sphere is $4\pi R^2$. The
magnetic length is $l_0 = \sqrt{\hbar c/|e| B}$.}
\label{fig3}
\end{figure}

In Fig. \ref{fig3} the density profiles for the exact ground state
$\psi_{gs} = \psi_{T1}(\lambda=0)$ (Exact), the trial state
$\psi_{T1}(\lambda=1)$ (VMC), the density obtained in FP DMC with trial
state $\psi_{T1}(\lambda=1)$ at time step $\tau=0.001$ (FP mixed
estimator), and the extrapolated density defined as ratio of the square
of FP density to the variational density corresponding to
$\psi_{T1}(\lambda=1)$, are shown. Note that since the density in our
DMC calculation is determined as a mixed estimate (see Eq.
\ref{mixede}), and the density operator does not commute with the
Hamiltonian between the DMC solution and the trial state, the
corresponding density profile (FP mixed estimator) improves on the
variational result but still differs from the exact density. The
extrapolated estimator for the density constructed by combining both,
the FP mixed estimator and the variational density makes it possible to
improve on the FP density and is seen to be very close to the exact
result.

{\it First Excited State} $(n=1)$: As a further demonstration of the
validity of our method we turn to the first excited state of the
electron-monopole system. Again, we specify the $L_{x^3}$ angular
momentum quantum number and take $m = {\cal S}+1$ for which the exact
excited state wave function is
\begin{eqnarray}
\psi_{es} = \left(\frac{|z|}{z(1+|z|^2)}\right)^{{\cal S}+1} |z| =
|\psi_{es}| e^{i\varphi_{es}}\ .
\end{eqnarray}
We then introduce two new trial states
\begin{equation}
\psi_{T1}=\left( \frac{|z|}{z (1 + |z|^2)} \right)^{{\cal S}+1}
\frac{|z|}{1+\lambda |z|^2} = |\psi_{T1}| e^{i\varphi_{T1}}
\end{equation}
and
\begin{equation}
\psi_{T2}=\left( \frac{|z|}{z (1 + |z|^2)} \right)^{{\cal S}+1} |z|
\left( \frac{|z|}{z} \right)^{\alpha} = |\psi_{T2}| e^{i\varphi_{T2}}
\end{equation}
with the property that for $\lambda=0$ and $\alpha = 0$ they each
reduce to the exact excited state, $\psi_{es}$.  As before, for
$\lambda \ne 0$ and $\alpha \ne 0$ the modulus of $\psi_{T1}$ is
approximate and the phase is exact ($|\psi_{T1}| \ne |\psi_{es}|$,
$\varphi_{T1} = \varphi_{es}$) and the modulus of $\psi_{T2}$ is exact
and the phase is approximate ($|\psi_{T2}| = |\psi_{es}|$,
$\varphi_{T2} \ne \varphi_{es}$).

\begin{figure}
\centerline{\psfig{figure=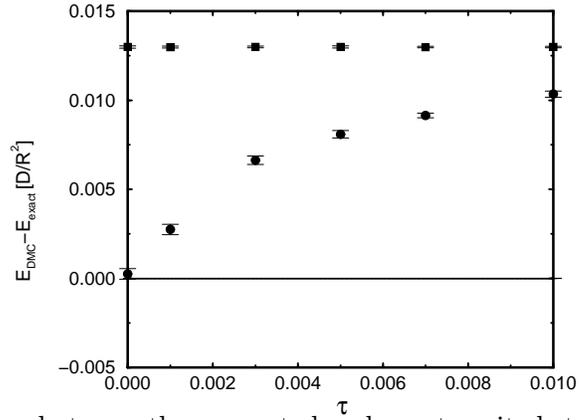,height=3in,angle=270}}
\caption{The difference between the computed and exact excited state
energies for the trial state with the exact phase $\psi_{T1}$
(circles) and the trial state with an approximate phase $\psi_{T2}$
(squares) for various values of time step. The extrapolated results to
time step $\tau = 0$ are also displayed.  For the trial state with the
exact phase one finds the exact solution and for the trial state with
an approximate phase one finds a variational upper bound for the exact
solution.}
\label{fig4}
\end{figure}

Figure~\ref{fig4} shows the difference between FP DMC energies
computed using trial states $\psi_{T1}$ and $\psi_{T2}$ and the exact
excited state energy for different values of time step $\tau$ as well
as the extrapolated $\tau = 0$ result.  The parameters (number of
walkers, number of Monte Carlo steps, etc.) used for these simulations
were the same as those used for the ground state simulations except
that we took $\alpha = 0.0015$ in $\psi_{T2}$. Again, our simulations
gave the expected results --- when the phase of the trial state is
exact we obtain the exact energy (circles), $E_1 = (3/2 + 1/{\cal S})
\hbar \omega_c$, and when the phase is approximate we obtain a
variational upper bound on that energy (squares).

\begin{figure}
\centerline{\psfig{figure=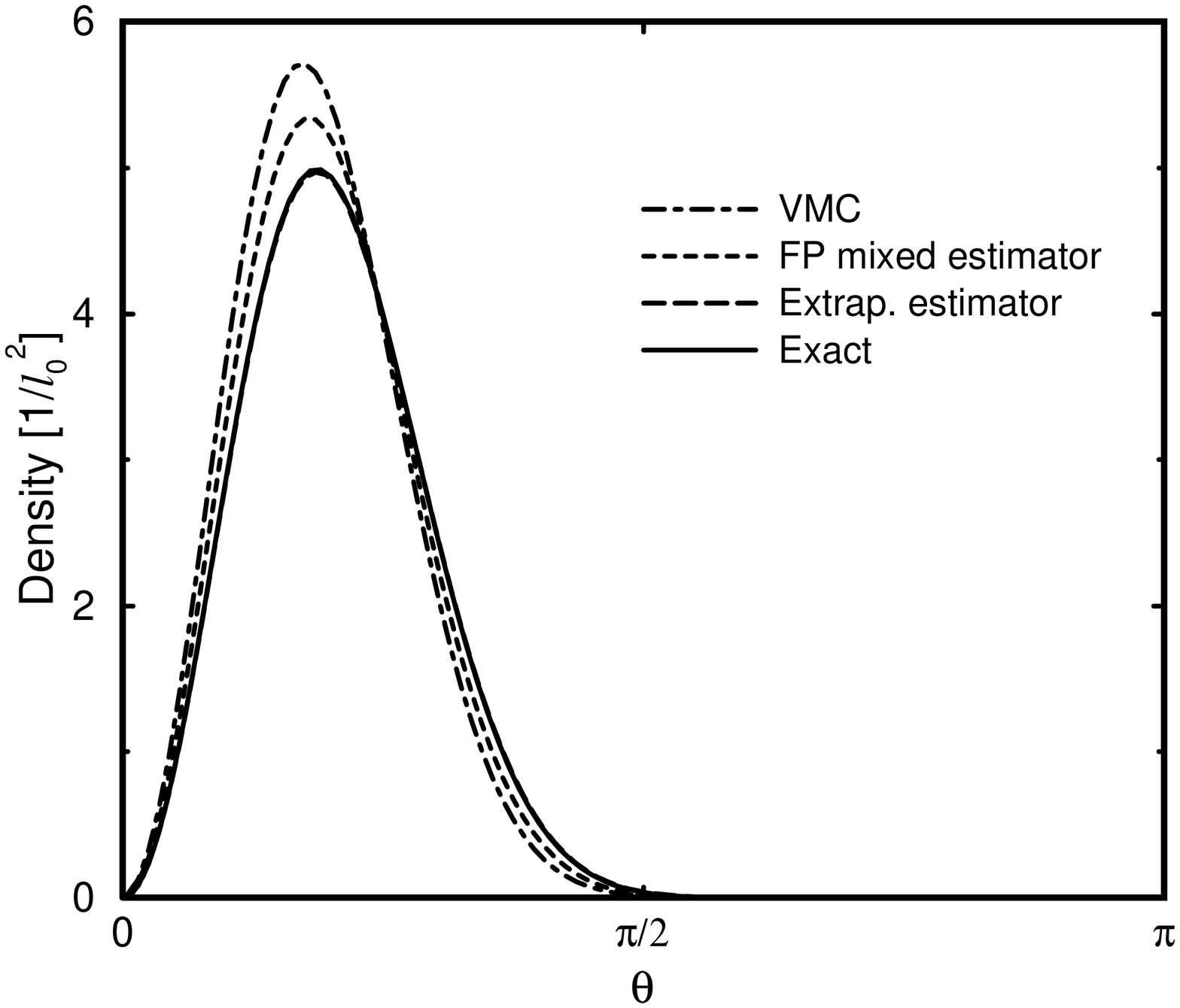,height=3in,angle=270}}
\caption{Excited state density corresponding to the trial state
$\psi_{T1}(\lambda = 1)$ (VMC), FP density with the same state at time
step $\tau = 0.001$ (FP mixed estimator), the extrapolated density,
which is computed by taking the ratio of square of the FP density and
the VMC density (Extrap. estimator), and the exact density (Exact). 
The FP estimator improves on the VMC result, but still differs from the
exact density. The extrapolated density allows one to improve on the FP
diffusion Monte Carlo result and is very close to the exact. The
density is normalized in such a way that its integral over the surface
of the sphere is $4\pi R^2$.}
\label{fig5}
\end{figure}

In Figure~\ref{fig5} the density profiles corresponding to the trial
state $\psi_{T1}(\lambda = 1)$ (VMC), the FP density using the same
state at time step $\tau = 0.001$ (FP mixed estimator), the
extrapolated density, computed as above by taking the ratio of square
of the FP density and the VMC density (Extrap. estimator), and the
exact density (Exact).  The results are qualitatively similar to those
for the ground state -- the FP estimator improves on the VMC result,
and the extrapolated density is nearly equal to the exact excited
state density.

The simulation results presented in this Section provide a simple test
of both the FP DMC method and the method developed in this paper for
dealing with quantum corrections due to curvature.  The results
clearly show that these methods can be used to study quantum systems
on curved manifolds.

\section{Discussion and conclusions}
\label{section6}

In this paper we have introduced a stochastic method to solve the
many-body Schr\"odinger equation on curved manifolds. This method is
essentially a generalized Diffusion Monte Carlo (DMC) technique
allowing one to deal with the effects arising from the space
curvature. We have shown that due to the curvature the diffusion
matrix and drift vector, which appear in the Green's function used as
a conditional probability in DMC simulations, acquire additional
terms, the so-called quantum corrections. The explicit expressions for
a general metric tensor are worked out in detail. Since the presence
of the curvature leads to a number of other nontrivial modifications,
we have presented a step by step algorithm which can be used to
implement a code dealing with DMC simulations in curved space.  

It is worth emphasizing that our method can be applied to a wide
variety of inhomogeneous systems (e.g., inhomogeneous semiconductors
with a position dependent effective mass), not just systems on curved
manifolds.  The reason for this, as discussed in Section III, is that
the quantum corrections to the Green's function can be interpreted as
being due to those terms which appear in the generalized diffusion
equation describing the system once each of the derivatives in that
equation have been commuted all the way to the left of each term.
This definition of quantum corrections is quite general and can be
applied to {\it any} differential equation which is second order in
space and first order in time, regardless of the number of dimensions
or any spatial inhomogeneity in the system.

To illustrate the general methodology we have concentrated on the
problem of interacting fermions in external electromagnetic
potentials. In this case a variational upper bound to the exact ground
state energy can be found by applying the Fixed-Phase approximation,
where the fermionic problem is treated as a bosonic one by fixing the
phase of the many-body wave function (which is complex-valued in
general) by some trial phase. As an example, we have considered the
problem of a single electron confined to the surface of a two-sphere,
which has a magnetic monopole at its center. The electron thus moves
in a space with curvature in the presence of a magnetic field which
breaks time-reversal symmetry. This simple problem can be solved in
closed form and, therefore, we have used it as a playground for
testing our technique. In the paper we have presented two
calculations, where the ground and first excited state energies are
computed using the exact phases, but approximate modulus for the
corresponding guiding functions. We have shown that the exact energies
are reproduced within statistical accuracy thus proving that the
approach for dealing with the quantum corrections is valid.

As emphasized in the Introduction, the method presented in this paper
for performing DMC simulations on curved manifolds can be used to
study many interesting physical systems.  An important example is the
quantum Hall effect, a phenomenon which occurs when a two-dimensional
electron system is placed in a strong magnetic field.  As first
pointed out by Haldane \cite{Haldane}, the electron-monopole system
described in Section V provides a convenient geometry for performing
finite size numerical studies of quantum Hall systems when many
interacting electrons are placed on the sphere. This is in part
because the spherical geometry has no boundary so that finite size
effects are suppressed. In addition, the spherical geometry is
conceptually simpler than the (flat metric) torus geometry, which also
has no boundary, because the topological order exhibited by quantum
Hall states leads to certain nontrivial degeneracies on the torus
\cite{Haldane-torus}.  

Recently we have used the method developed in this paper to study some
of the exotic excitations which occur in quantum Hall systems,
specifically the fractionally charged quasiparticle excitations of the
fractional quantum Hall effect \cite{gaps}, and the charged spin
texture excitations (skyrmions) of the integer quantum Hall effect
\cite{skyrmion}. Previous numerical studies of these excitations have
been based on either VMC or exact diagonalization calculations which,
for the most part, have assumed that the wave functions describing the
excitations are confined to the lowest ($n=0$) Landau level.  In fact,
this is a rather poor approximation for real experimental systems
which can exhibit significant mixing of higher Landau levels due to
the electron-electron interaction.  Because the FP DMC method allows
one to go beyond the lowest Landau level approximation it can be used
to study the effect of Landau level mixing on quantum Hall states
\cite{fp} and, by employing the method described in this paper, we
were able to perform such studies using Haldane's spherical geometry.
These simulations have provided useful quantitative results for
various properties of quantum Hall excitations for realistic
experimental parameters \cite{gaps,skyrmion}.  Along with the more
rigorous test case of the electron-monopole system described in this
paper, these calculations of Landau level mixing effects in quantum
Hall systems using the Haldane sphere have shown that the method we
have developed for performing FP DMC calculations on curved manifolds
works well and can be used to study many other interesting physical
systems.

\acknowledgements

We acknowledge stimulating discussions with David Ceperley. VM is
supported by NSF grant number DMR-9725080. NEB is supported by US DOE
grant number DE-FG02-97ER45639.  Work at Los Alamos is sponsored by
the US DOE under contract W-7405-ENG-36.




\newpage

\newpage

\newpage

\newpage

\newpage



\begin{references}

\bibitem{mattis} D.C. Mattis, {\it The Many-Body Problem} (World
Scientific, Singapore, 1993).

\bibitem{negele}
For an introduction see, e.g., J.W. Negele and H. Orland, {\it Quantum
Many-Particle Systems} (Addison Wesley, Redwood City, 1988), Ch. 8.

\bibitem{anderson}
P.W. Anderson, {\it Basic Notions of Condensed Matter Physics} (Addison
Wesley,  Reading, 1984).

\bibitem{ceper}
D.M. Ceperley, in {\it Recent Progress in Many-Body Theories}, edited by
E. Schachinger, H. Mitter, and M. Sormann (Plenum, 1995).

\bibitem{gravitation}
L. Parker, Phys.  Rev. D {\bf 22}, 1922 (1980).

\bibitem{curve} 
C. Yannouleas, E. N. Bogachek, and U. Landman, Phys.  Rev. B {\bf 53},
10225 (1996).

\bibitem{fp} 
G. Ortiz, D.M. Ceperley, and R.M. Martin, Phys. Rev.  Lett. {\bf 71},
2777 (1993).

\bibitem{landau}
L. Landau, Z. Phys. {\bf 64}, 629 (1930). 

\bibitem{hofst} 
C. Kreft and R. Seiler, J. Math. Phys. {\bf 37}, 5207  (1996). 

\bibitem{QHE} 
{\it Perspectives in Quantum Hall Effects}, edited by  S. Das Sarma and
A. Pinczuk (Wiley, New York, 1997). 

\bibitem{tira} 
F. Langouche, D. Roekaerts, and E. Tirapegui, Il Nuovo  Cimento {\bf 53
B}, 135 (1979).  

\bibitem{landauctf}
L.D. Landau and E.M. Lifshitz, {\it The Classical Theory of Fields}
(Butterworth-Heinemann, Oxford, 1999), Ch. 10.

\bibitem{hammer} 
M. Hamermesh, {\it Group Theory} (Addison Wesley,  Reading, 1962). 
 
\bibitem{qc}
For a discussion of quantum corrections see, for example, B.S. DeWitt,
Rev. Mod. Phys. {\bf 29}, 377 (1957); and D.W. McLaughlin and L.S.
Schulman, J. Math. Phys. {\bf 12} 2520 (1971).

\bibitem{polya} 
A. M. Polyakov, {\it Gauge Fields and Strings}  (Harwood, Chur, 1993),
p.176. 

\bibitem{feynman} 
R. P. Feynman and A. R. Hibbs, {\it Quantum  Mechanics and Path
Integrals} (Mc Graw-Hill, New York, 1965), Ch.4.

\bibitem{onsager} 
L. Onsager and S. Machlup, Phys. Rev. {\bf 91}, 1505  (1953). 
  
\bibitem{Note1} 
For real symmetric Hamiltonians, this catastrophe is known as the
``fermion-sign'' problem. For a recent study, see M.H. Kalos and K.E.
Schmidt, J. Stat. Phys. {\bf 89}, 425 (1997). 
 
\bibitem{Wilczek} 
F. Wilczek, {\it Fractional Statistics and Anyon Superconductivity}
(World Scientific, Singapore, 1990).

\bibitem{vortex} 
G. Ortiz and D.M. Ceperley, Phys. Rev. Lett. {\bf 75}, 4642 (1995). 

\bibitem{reynolds} 
P.J. Reynolds, D.M. Ceperley, B.J. Alder, and W.A.  Lester, Jr., J.
Chem. Phys. {\bf 77}, 5593 (1982).

\bibitem{Haldane}
F.D.M. Haldane, Phys. Rev. Lett. {\bf 51}, 605 (1983).

\bibitem{Wu}
T.T. Wu and C.N. Yang, Nucl. Phys. B {\bf 107}, 365 (1976).

\bibitem{Haldane-torus} F.D.M. Haldane, Phys. Rev. Lett. {\bf 61},
1029 (1988).


\bibitem{gaps}
V. Melik-Alaverdian, N.E. Bonesteel, and G. Ortiz, Phys. Rev.  Lett. 
{\bf 79}, 5286 (1997).

\bibitem{skyrmion}
V. Melik-Alaverdian, N.E. Bonesteel, and G. Ortiz, Phys. Rev. B 
{\bf 60}, R8501 (1999).


\end{references}
\end{document}